\documentclass[12pt]{spieman}  

\usepackage{setspace}
\usepackage{tocloft}

\newcommand{\LP}[1]{${\rm LP_{#1}}$}
\usepackage{amsmath,amsfonts,amssymb}
\usepackage{graphicx}
\usepackage[colorlinks=true, allcolors=blue]{hyperref}

\title{Spectral characterization of a 3-port photonic lantern for application to spectroastrometry}

\author[a]{Yoo Jung Kim}
\author[a]{Michael P. Fitzgerald}
\author[a]{Jonathan Lin}
\author[b]{Julien Lozi}
\author[b,c]{S\'ebastien Vievard}
\author[d]{Yinzi Xin}
\author[e]{Daniel Levinstein}
\author[d]{Nemanja Jovanovic}
\author[f]{Sergio Leon-Saval}
\author[f]{Christopher Betters}
\author[b,c,g,h]{Olivier Guyon}
\author[f]{Barnaby Norris}
\author[e]{Steph Sallum}

\affil[a]{Department of Physics \& Astronomy, 430 Portola Plaza, University of California, Los Angeles, CA 90095, USA}
\affil[b]{Subaru Telescope, National Observatory of Japan, HI 96720, USA}
\affil[c]{Astrobiology Center, 2-21-1, Osawa, Mitaka, Tokyo, 181-8588, Japan}
\affil[d]{Department of Astronomy, California Institute of Technology, Pasadena, CA 91125, USA}
\affil[e]{Department of Physics \& Astronomy, University of California Irvine, 4129 Frederick Reines Hall, Irvine, CA 92697, USA}
\affil[f]{Sydney Astrophotonic Instrumentation Laboratory, School of Physics, The University of Sydney, Sydney, NSW 2006, Australia}
\affil[g]{Steward Observatory, University of Arizona, Tucson, AZ 85721, USA}
\affil[h]{College of Optical Sciences, University of Arizona, Tucson, AZ 85721, U.S.A.}

\graphicspath{{./}{figures/}}

\usepackage{lineno}

\begin{document} 
\maketitle

\providecommand\mnras{Monthly Notices of the Royal Astronomical Society}
\providecommand\apj{The Astrophysical Journal}
\providecommand\apjl{The Astrophysical Journal, Letters}
\providecommand\apss{Astrophysics and Space Science}
\providecommand\aap{Astronomy \& Astrophysics}
\providecommand\pasp{Publications of the Astronomical Society of the Pacific}
\providecommand\nat{Nature}

\begin{abstract}
Spectroastrometry, which measures wavelength-dependent shifts in the center of light, is well-suited for studying objects whose morphology changes with wavelength at very high angular resolutions. Photonic lantern (PL)-fed spectrometers have potential to enable measurement of spectroastrometric signals because the relative intensities between the PL output SMFs contain spatial information on the input scene. In order to use PL output spectra for spectroastrometric measurements, it is important to understand the wavelength-dependent behaviors of PL outputs and develop methods to calibrate the effects of time-varying wavefront errors in ground-based observations. We present experimental characterizations of the 3-port PL on the SCExAO testbed at the Subaru Telescope. We develop spectral response models of the PL and verify the behaviors with lab experiments. We find sinusoidal behavior of astrometric sensitivity of the 3-port PL as a function of wavelength, as expected from numerical simulations. Furthermore, we compare experimental and numerically simulated coupling maps and discuss their potential use for offsetting pointing errors. We then present a method of building PL spectral response models (solving for the transfer matrices as a function of wavelength) using coupling maps, which can be used for further calibration strategies.
\end{abstract}


\section{Introduction}

The ability to achieve high angular resolution in the visible and near infrared is crucial in many fields in astronomy, such as for studies of the innermost regions of protoplanetary disks and broad-line regions of active galactic nuclei. Combined with high resolution spectroscopy, spectroastrometry is a powerful technique of studying kinematics and morphology at small angular scales. Spectroastrometry is a method of studying angular scales smaller than the resolution limit, by measuring the relative position of an extended but unresolved object as a function of wavelength \cite{bailey98, whelan08}. If a spectral line originating from an extended object is broadened due to its kinematic structures such as rotation, the kinematic structures can be probed using spectroastrometry on the line. With diffraction-limited point-spread functions (PSFs) enabled by adaptive optics (AO), sub-milliarcsecond precisions can be achieved with 10\,m-class telescopes in the near infrared. Combined with long-baseline interferometry, even better precision can be achieved \cite{gravity_yso_i}. Indeed, spectroastrometry has been applied to a number of studies probing the innermost regions of protoplanetary disks and broad-line regions (e.g., \cite{pontoppidan08, pontoppidan11, brittain15, bosco21, gravity18, gravity_yso_i, gravity_yso_vii, gravity_yso_x}).  

However, achieving both high angular resolution and high spectral resolution is demanding. Long-slit echelle spectrometers have been typically used for spectroastrometric observations, since they can achieve high spectral resolution and spatial sampling along one axis. With long-slit spectrometers, however, at least two observations with different slit orientations are required in order to obtain two-dimensional spectroastrometric signals. Another downside of using a slit is that a distorted PSF or uneven illumination can introduce an artificial spectroastrometric signal \cite{bra06, whelan15}. At medium resolutions, integral field spectrometers can enable two-dimensional spectroastrometry with decreased artifacts, although with an increased complexity \cite{dav10, got12, mur13}. Spectro-interferometry can enable both high angular resolution and spectral resolution by combining beams from separate apertures and measuring differential visibility signals as a function of wavelength. Nevertheless, it necessitates significantly more complex setups, either involving pupil remapping within a single telescope or utilizing multiple telescopes.

\begin{figure}
    \centering
    \includegraphics[width=1\linewidth]{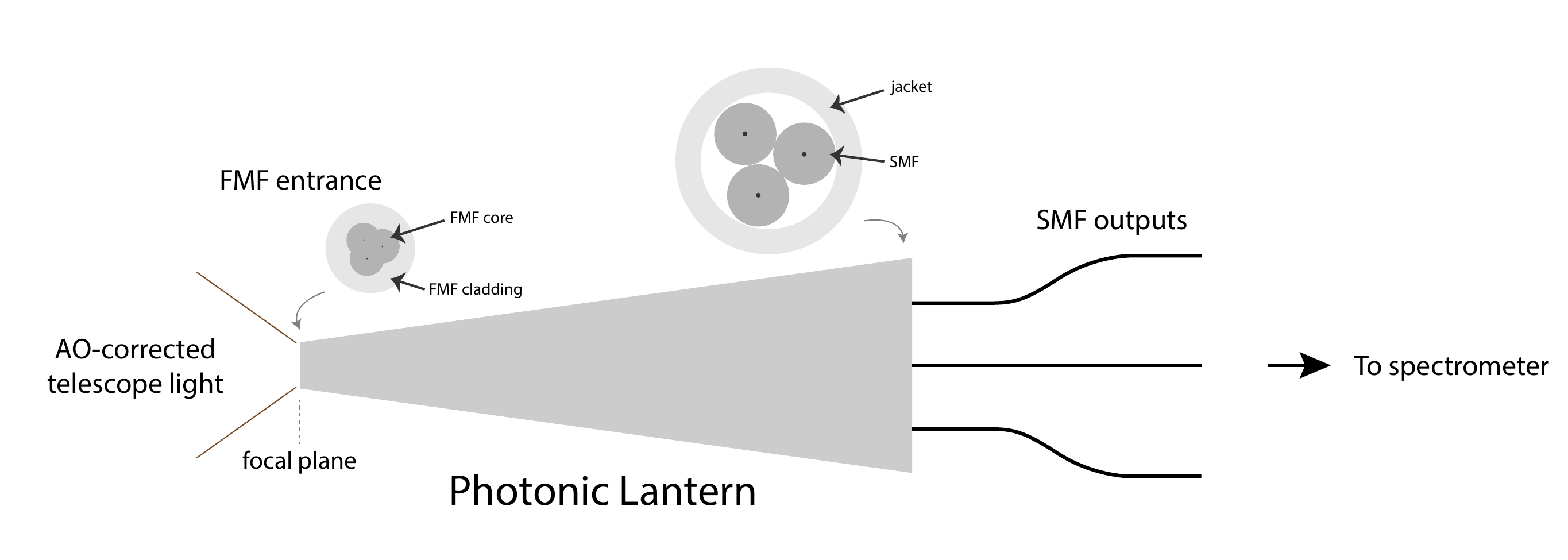}
    \caption{A schematic diagram of the standard 3 port photonic lantern (PL) and its cross sections in the two ends. The telescope light is coupled in the FMF entrance, propagates through the lantern, and becomes effectively confined in the SMFs. The light coupled in the SMFs can be fed to a spectrometer for spectroscopy. }
    \label{fig:PL}
\end{figure}

Ref \cite{kim24JATIS} (hereafter Paper I) studies the concept of using few-moded photonic lanterns (PLs) for spectroastrometry, which can simultaneously achieve high angular resolution and high spectral resolution. The PL is a tapered waveguide that gradually transitions from a few-mode fiber (FMF) geometry to a bundle of single-mode fibers (SMFs) \cite{leo10, leo13, bir15}. Its FMF end can be placed in the telescope focal plane where AO-corrected telescope light can couples into it. The light becomes confined within the SMF cores as it propagates through the lantern (Figure \ref{fig:PL}). Thus, a PL converts AO-corrected few-moded telescope light into multiple single-moded beams. When the input beam moves around or changes the shape due to aberrations, the distribution of light amongst the SMF outputs changes. Due to the dependence of the PL outputs on the input wavefront, PLs have recently been studied and demonstrated for focal plane low-order wavefront sensor applications (PLWFS; \cite{cor18, nor20, cruz-delgado21, lin22, lin23, lin23b}). When the SMF outputs are spectrally dispersed, relative intensities as a function of wavelength represent spectroastrometric signals. The SMF outputs of a few-moded PL are suitable for feeding high spectral resolution diffraction-limited spectrometers \cite{lin21, vievard24}, allowing both high angular and high spectral resolution measurements.

In this study, we present an experimental characterization of the near-infrared 3-port PL on the SCExAO testbed \cite{jovanovic15} for spectroastrometric measurements. A monochromatic characterization of a near-infrared 19-port PL on the SCExAO testbed has been done by Ref. \cite{lin22spie} which focused on studying the overall coupling efficiency as a function of injection parameters. 
Ref. \cite{vievard24} studied the behavior of the summed spectra and the overall coupling efficiency of the visible 19-port PL feeding the FIRST spectrograph on the SCExAO visible bench.
In this work, we focus on studying the dispersed PL outputs of individual ports, specifically the wavelength dependence of the PL's sensitivity to aberrations and thus spectroastrometric signals (Paper I). An overview of the concept of PL spectroastrometry is detailed in the next section.

\section{Spectroastrometry with PLs}\label{sec:SA}

Paper I presents the concept of PL spectroastrometry, including numerically simulated spectroastrometric signals for simple astronomical scenes, effects of photon noise, wavefront errors (WFEs), and chromatic behavior of the PLs. We provide a brief summary of the concept in this section.

PL output intensities are sensitive to low-order structures in the input scene. The pupil plane tip-tilt modes, the lowest order pupil plane aberration, correspond to light centroid displacement in the focal plane. The goal of PL spectroastrometry is to measure the light centroid shifts as a function of wavelength, which relies on the sensitivity of PL output normalized intensities (output intensities divided by the intensity sum of the all the outputs) to tip-tilt.
In the small tip-tilt regime, for an $N$-port PL, the (monochromatic) intensity response of the PL outputs can be linearized as 
\begin{equation}\label{eq:linear}
    \textbf{I}_{{n}} \approx \textbf{I}_{n0} + B_n \boldsymbol{\alpha}_{\rm centroid}
\end{equation}
where $\textbf{I}_{{n}}$ is the array of $N$ normalized intensities, $\textbf{I}_{{n0}}$ is the array of $N$ normalized intensities for a point source without centroid shifts (reference signal), $B_n$ is the tip-tilt ($x$,$y$) linear (first-order) normalized intensity response matrix ($N \times 2$ dimensional), and $\boldsymbol{\alpha}_{\rm centroid}$ is the array describing tip-tilt ($x$,$y$ centroid of the input scene), measured in radians with zero for on-axis. Thus, by measuring the normalized intensity $\textbf{I}_{{n}}$ and with the experimentally characterized linear intensity response $B_n$ and $\textbf{I}_{{n0}}$, the centroid shift $\boldsymbol{\alpha}_{\rm centroid}$ can be determined.

The $B_n$ matrix is the important factor for spectroastrometric measurements, describing the sensitivity to astrometric displacements in the small tip-tilt regime. Paper I presents that the $B_n$ matrix is in fact a function of wavelength, $B_n = B_n(\lambda)$. 
Unless the PL is designed to be achromatic (such as a mode-selective PL \cite{xin24}), the $B_n$ matrix is expected to depend on wavelength due to propagation constant differences between the modes propagating along the PL. For a standard 3-port PL, the $B_n$ matrix is expected to have sinusoidal dependence on wavelength, which means that the astrometric sensitivity of the PL oscillates with wavelength. Paper I also studies the behavior of the $B_n$ matrix under static and time-varying WFEs with numerical simulations. The static WFE such as a small systematic tip-tilt changes mostly the reference signal $\textbf{I}_{{n0}}$, but can also affect the $B_n$ matrix for larger aberrations. The time-varying WFE averaged over an exposure modifies the $B_n$ matrix, degrading the astrometric sensitivity.

The focus of this paper is the experimental characterization of the spectral behavior of $\textbf{I}_{{n0}}$ and $B_n$ of a standard 3-port PL. In \S\ref{sec:model}, we first describe a general analytic model of PL spectra. In \S\ref{sec:experiment}, we present our experimental results on characterizing the wavelength dependence of $B_n$ (\S\ref{ssec:tiptilt}) and its behavior with time-varying WFEs (\S\ref{ssec:tiptilt_WFE}). We also present coupling maps in individual ports as a function of wavelength, which can be used to correct for systematic tip-tilt, or pointing errors (\S\ref{ssec:couplingmap}). Moreover, in \S\ref{ssec:mapfitting} we show the preliminary attempt on building PL spectral response models based on \S\ref{sec:model} that could be used for calibrating any misalignments and effects of WFEs. In \S\ref{sec:conclusion} we summarize our results and discuss directions for future work.

\section{Model of PL spectra}\label{sec:model}

In this section, we describe a model of the PL output spectra, given an input scene. We first consider the monochromatic case in \S\ref{ssec:monochromatic} and introduce a method to decompose PL output intensities into linear combinations of overlap integrals between the fiber modes and the input field using transfer matrices. In \S\ref{ssec:Bn} we provide the connection between the model and the intensity response to tip-tilt described in \S\ref{sec:SA}. In \S\ref{ssec:chromaticity}, we discuss the wavelength dependence of the PL transfer matrix and its implications on wavelength-dependent angular resolutions. In \S\ref{ssec:WFE} we describe the effects of averaging intensities in presence of time-varying phase aberrations. 

\subsection{Model of PL output intensities in terms of the PL transfer matrix}\label{ssec:monochromatic}

The (monochromatic) complex amplitude in the $i$-th SMF output of a PL can be expressed as an overlap integral over the pupil plane
\begin{equation}\label{eq:E_i}
    \tilde{E}_i = \int \tilde{E}_p \tilde{P}_{{\rm eff},i} dA
\end{equation}
where $\tilde{E}_p$ represents the pupil plane electric field and $\tilde{P}_{{\rm eff},i}$ denotes the effective pupil function of the $i$-th output, which is the telescope aperture function multiplied by the complex conjugate of the $i$-th pupil plane PL principal mode (PLPM; \cite{kim24}). We use ($\tilde{.}$) to indicate complex-valued functions. The pupil plane PLPM ($\tilde{P}_{p,i}$) is the inverse Fourier transform of the focal plane PLPM ($\tilde{P}_{f,i}$; at the fiber entrance). The focal plane PLPM is an orthogonal superposition of the guided entrance modes of a PL (scalar LP modes if assuming a circular symmetric step-index fiber) that maps to one SMF output. Assuming that PLPMs consist of a pure linear combination of PL entrance modes and that the number of guided modes is equal to the number of ports ($N$), the $i$-th focal plane PLPM can be decomposed as
\begin{equation}\label{eq:Pi_focal}
    \tilde{P}_{f,i} = \sum_{j=1}^{N} \tilde{\eta}_{ij} {X}_{f,j}.
\end{equation}
${X}_{f,j}$ is the $j$-th FMF eigenmode (real-valued fiber entrance scalar spatial mode) under the weakly guiding approximation and $\tilde{\eta}_{ij}$ represents the $i,j$ component of the PL transfer matrix that describes how focal plane fiber entrance modes are mapped to individual ports \cite{lin23}. In the pupil plane, the pupil plane PLPM (the inverse Fourier transform of the focal plane PLPM) can be written as
\begin{equation}
    \tilde{P}_{p,i} = \sum_{j=1}^{N} \tilde{\eta}_{ij} \tilde{X}_{p,j}
\end{equation}
where $\tilde{X}_{p,j}$ is the inverse Fourier transform of the $j$-th fiber entrance mode $X_{f,j}$. $\tilde{\eta}_{ij}$ can be written in a phasor form, $\tilde{\eta}_{ij} = \eta_{ij} \exp{(i\phi_{ij})}$, with real and positive $\eta_{ij}$ and $\phi_{ij}$ in [0,$2\pi$).

Using the decomposed expression, Equation \ref{eq:E_i} can be expressed as 
\begin{equation}
    \tilde{E}_i = \sum_{j=1}^{N} \tilde{\eta}_{ij}^* \int \tilde{E}_p \tilde{X}_{p,j}^* T dA.
\end{equation}
with $T$ being the telescope aperture transmission function. The integral is evaluated over the pupil plane. The intensity in port $i$ can be expressed as
\begin{align}
\begin{aligned}\label{eq:intensity1}
    {I}_i = |\tilde{E}_i|^2 &= \sum_{j=1}^{N} \eta_{ij}^2 {\rm Re}{(\tilde{W}_{jj})}  +\sum_{j=k+1}^{N}\sum_{k=1}^{N} 2\eta_{ij}\eta_{ik}  \left(\cos{(\phi_{ik}-\phi_{ij})}~ {\rm Re}{(\tilde{W}_{jk})} - \sin{(\phi_{ik}-\phi_{ij})} ~{\rm Im}{(\tilde{W}_{jk})}\right)\\
    &= \sum_{j=1}^{N} \eta_{ij}^2 {\rm Re}{(\tilde{W}_{jj})}  +\sum_{j=k+1}^{N}\sum_{k=1}^{N} 2\eta_{ij}\eta_{ik} |\tilde{W}_{jk}| \cos{(\phi_{ik}-\phi_{ij} + \arg{(\tilde{W}_{jk})})}
\end{aligned}
\end{align}
where 
\begin{equation}\label{eq:overlap}
    \tilde{W}_{jk} \equiv \iint{\tilde{J}_p(u,v)~ \left[(\tilde{X}_{p,j} T) \star (\tilde{X}_{p,k} T)\right](u,v)~ dudv}. 
\end{equation}
Here, $\tilde{J}_p (u,v)$ is the mutual intensity in the pupil plane, $\tilde{J}_p (u,v) = \tilde{E}_p(x,y) \tilde{E}^*_p(x+\lambda u, y+\lambda v)$, a measure of spatial coherence between pupil plane points ($x$, $y$) and ($x + \lambda u$, $y + \lambda v$) where $\lambda$ is the wavelength and ($u,v$) denotes two-dimensional spatial frequency. Assuming that the source field is spatially incoherent, $\tilde{J}_p (u,v)$ corresponds to the Fourier transform of the source intensity distribution by van Cittert-Zernike theorem. $\star$ denotes cross-correlation. The integral is over the entire two-dimensional frequency plane. 

\begin{figure}[hbt!]
    \centering
    \includegraphics[width=0.7\linewidth]{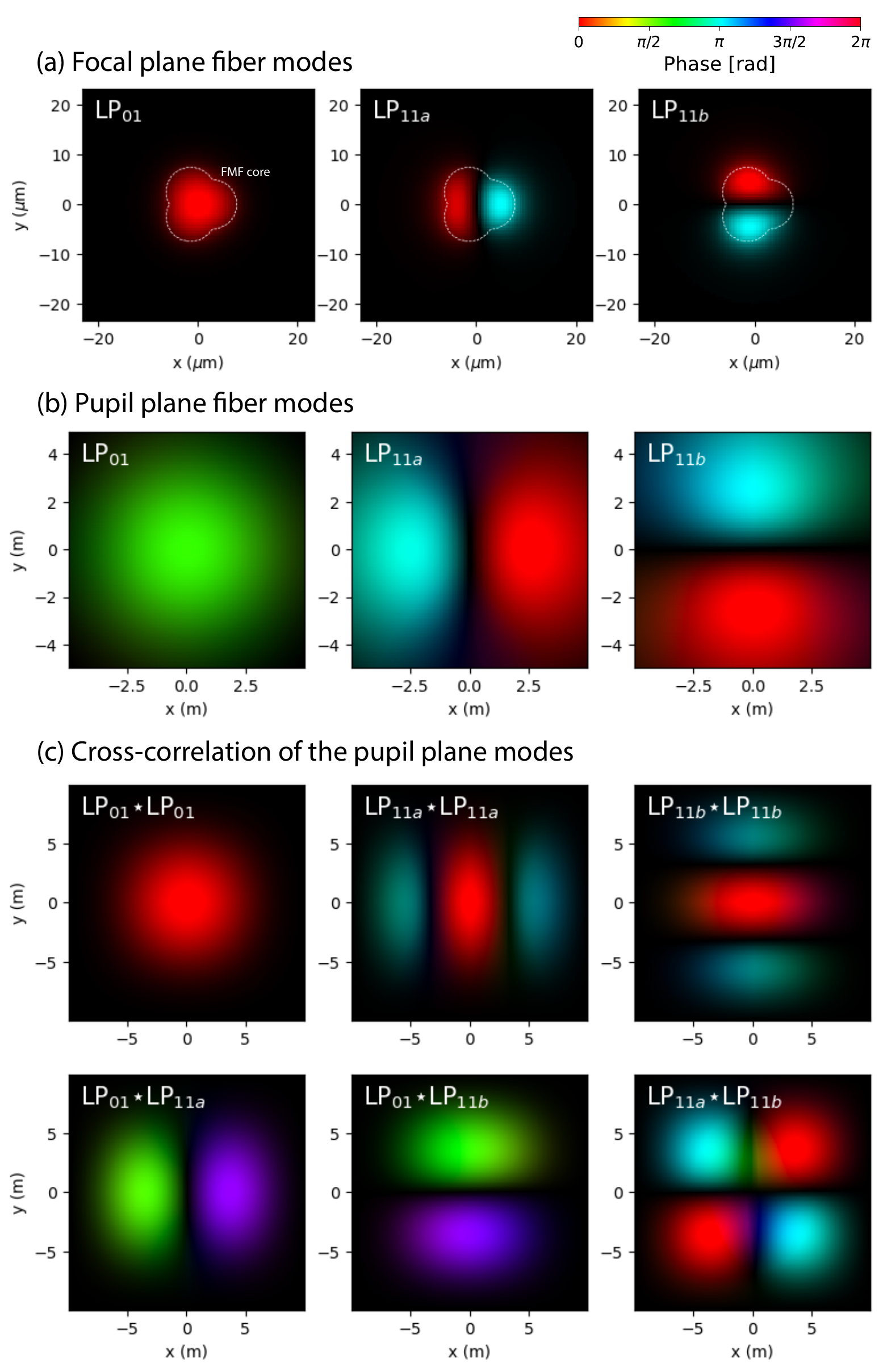}
    \caption{(a) Focal plane fiber entrance modes of the flower-shaped 3-port PL, of which the geometry is matched to the 3-port PL on SCExAO. The white dotted lines indicate the interface between the FMF core and the cladding (or SMF cladding-jacket interface). The modes are close to the \LP{01}, \LP{11a}, \LP{11b} modes but with small asymmetries. (b) Pupil plane fiber modes, computed by Fourier transform of the modes in (a). Note that pupil plane \LP{01} mode is nearly purely imaginary since the focal plane \LP{01} mode is nearly even. Pupil plane \LP{11} modes are nearly purely real since the focal plane \LP{11} modes are nearly odd. (c) The cross-correlation of the pupil plane modes, which are frequency sampling functions of $\tilde{J}_p$.}
    \label{fig:modes}
\end{figure}

The correspondence of the cross-correlation of pupil functions to the optical transfer functions (OTFs) was discussed in Ref. \cite{kim24}. The intensity in each SMF output corresponds to the pupil plane mutual intensity function ($\tilde{J}_p$) weighted by the autocorrelation of the pupil function integrated over all the frequencies, which can then be decomposed into $N(N+1)/2$ components as in Equation \ref{eq:intensity1}. Note that in Equation \ref{eq:intensity1}, all the dependence on the input scene is isolated in the $\tilde{W}_{jk}$ term and is common to all the ports. The transfer matrix amplitude components $\eta_{ij}$s determine the relative importance of the terms. The relative phases $\phi_{ik} - \phi_{ij}$ also play an important role in determination of the relative contribution of the real and imaginary parts of $\tilde{W}_{jk}$ in the intensities. Such decomposition is useful because the relative phases change with wavelength and determine the PLs' wavelength-dependent sensitivity to aberrations, as will be discussed in \S\ref{ssec:chromaticity}.

In Figure \ref{fig:modes} we visualize the modes and cross-correlation of them for an example 3-port PL. Panel (a) shows the fiber entrance modes, ${X}_{f,j}$. The supported LP modes for a 3-moded circular step-index fiber are \LP{01}, \LP{11a}, and \LP{11b} modes; however, a realistic 3-port PL may not exhibit circular symmetry, due to the flower-shaped core geometry formed by fusing three SMFs arranged in an equilateral triangle, as shown in Figure \ref{fig:PL}. We solve ${X}_{f,j}$s for the geometry of the 3-port PL on SCExAO (see \S\ref{ssec:setup}) using {\tt wavesolve}.\footnote{\url{https://github.com/jw-lin/wavesolve}} The SMF residual cores in the PL entrance are ignored since their effects are negligible. Panel (b) shows the pupil plane modes ($\tilde{X}_{p,j}$), computed by the inverse Fourier transform of the fiber entrance modes. Panel (c) shows the cross-correlation of every pair of the pupil plane modes, $\tilde{X}_{p,j} \star \tilde{X}_{p,k}$. 

\subsection{Intensity responses to small tip-tilt (astrometric signals)}\label{ssec:Bn}

Now let us relate the models to the linear intensity response to tip-tilt ($B_n$), for spectroastrometry. 
From Equation \ref{eq:intensity1}, we can gain insight on how the intensities change upon small tip-tilt aberration.

For tilt displacement of $\theta$ in radians along $x$, for instance, the input field can be written as $\tilde{E}_p(x) = \exp{(i2\pi\theta x/\lambda)}$, and the mutual intensity as  $\tilde{J}_p(u,v) = \exp{(-i2\pi \theta u)} = \cos{(2\pi \theta u)} - i\sin{(2\pi \theta u)}$. For the case of 3-port PL and for small angular scales $\theta$, the frequency overlap of \LP{01} $\star$ \LP{11a} with the sinusoidal imaginary part of the $\tilde{J}_p(u,v)$ mainly contributes to the intensity response of the PL on the tilt displacement. Since \LP{01} $\star$ \LP{11a} is nearly purely imaginary-valued (see Figure \ref{fig:modes}), Equation \ref{eq:overlap} implies that the intensity response to tilt is in the real part of $\tilde{W}_{\rm LP_{01}, LP_{11a}}$. Thus, the coefficient $2\eta_{i,{\rm LP_{01}}} \eta_{i,{\rm LP_{11a}}} \cos{(\phi_{i,{\rm LP_{11a}}}- \phi_{i,{\rm LP_{01}}})}$ determines the amplitude of the sensitivity to the tilt along $x$. Likewise, the coefficient $2\eta_{i,{\rm LP_{01}}} \eta_{i,{\rm LP_{11b}}} \cos{(\phi_{i,{\rm LP_{11b}}}- \phi_{i,{\rm LP_{01}}})}$ determines the amplitude of the sensitivity to the tilt along $y$.
Relating this to the $B_n$ matrix introduced in \S\ref{sec:SA}, the $x$, $y$ component of the $B_n$ matrix (spatial frequency along $u$, $v$) in port $i$ can be written as
\begin{align}
\begin{aligned}\label{eq:Bn_3port}
    B_{n, ix} &\approx \frac{1}{I_{\rm sum}} \left( 2 \eta_{i,\rm LP_{01}}\eta_{i,{\rm LP_{11a}}} \cos{(\phi_{i,{\rm LP_{11a}}} - \phi_{i,{\rm LP_{01}}})} \iint{2\pi u ~{\rm Im}\left[(\tilde{X}_{p,{\rm LP_{01}}} T) \star (\tilde{X}_{p,{\rm LP_{11a}}} T)\right](u,v)~ dudv} \right) \\
    B_{n, iy} &\approx \frac{1}{I_{\rm sum}} \left( 2  \eta_{i,\rm LP_{01}}\eta_{i,{\rm LP_{11b}}} \cos{(\phi_{i,{\rm LP_{11b}}} - \phi_{i,{\rm LP_{01}}})} \iint{2\pi v ~{\rm Im}\left[(\tilde{X}_{p,{\rm LP_{01}}} T) \star (\tilde{X}_{p,{\rm LP_{11b}}} T)\right](u,v)~ dudv} \right) 
\end{aligned}
\end{align}
where $I_{\rm sum}$ corresponds to the intensity sum over all the ports, $I_{\rm sum} = \sum_{i} I_i$.

\subsection{Wavelength dependence of the PL transfer matrix}\label{ssec:chromaticity}

Paper I reports on a study of the chromaticity of PLPMs with numerical simulations of standard 3-port and 6-port PLs. Due to propagation constant differences between the spatial supermodes, relative phases between the fiber modes that constitute PLPMs slowly vary with wavelength. For the 3-port PL case, the relative phase between \LP{01} and \LP{11} modes varies nearly linearly as a function of wavelength.
Consequently, the intensity response to tip-tilt depends on wavelength, in a sinusoidal manner, due to the $\cos{(\phi_{i,{\rm LP_{11}}}- \phi_{i,{\rm LP_{01}}})}$ term (Equation \ref{eq:Bn_3port}). When the phase difference between \LP{01} and the \LP{11} modes in the $i$-th port is a multiple of $\pi$, the 
absolute values of $B_n$ matrix of
the $i$-th port (the linear intensity response) are maximized, implying large sensitivity to small angular resolutions. Conversely, when the phase difference is an odd multiple of $\pi/2$, the intensity response of the $i$-th port is minimized. 
See Figure 9 of Paper I for details.
In \S\ref{ssec:tiptilt}, we verify this behavior with experiments.
 
\subsection{Effects of WFE averaging}\label{ssec:WFE}

Now we consider the effects of time-varying phase aberrations in a long exposure regime. The pupil plane phase aberration $\varphi$ modifies the telescope aperture function $T(x,y)$ to $\tilde{T}(x,y) = T(x,y) \exp{(i\varphi(x,y))}$. Let us define the time-averaged OTF of the phase screen as $\langle \Gamma_{\rm turb} (u,v)\rangle_t$. The subscript $t$ refers to time average. In the long exposure regime, we can write the Equation \ref{eq:overlap} as
\begin{align}
\begin{aligned}
    \langle \tilde{W}_{jk, {\rm turb}}\rangle_t 
    &\approx \iint{\tilde{J}_p(u,v)~ \left[(\tilde{X}_{p,j} T) \star (\tilde{X}_{p,k} T)\right](u,v)~ \langle \Gamma_{\rm turb} (u,v) \rangle_t ~dudv}.
\end{aligned}
\end{align}
Note that $\langle \Gamma_{\rm turb} \rangle_t$ can be related to the phase structure function, $\exp{(-\frac{1}{2} \langle|\varphi(\vec{r}) - \varphi(\vec{r}+\vec{\delta r})|^2\rangle)} = \exp{(-\frac{1}{2}D_\varphi)}$, with negligible imaginary part \cite{roddier81}. Due to this component, higher frequencies in the frequency sampling functions $(\tilde{X}_{p,j} T) \star (\tilde{X}_{p,k} T)$ are suppressed. 

The WFE-averaged $B_n$ matrix, $\langle B_n \rangle_t$, is modified in a similar way. 
For the case of 3-port PL, the same $\langle \Gamma_{\rm turb} (u,v)\rangle_t$ factor goes into the integrals of Equation \ref{eq:Bn_3port}. The summed intensity $I_{\rm sum}$ also decreases with WFE averaging, but not as fast as the numerator, as the reduction in coupling into the \LP{01} mode is partially compensated by the increased coupling into \LP{11} modes. These result in a systematic decrease in the intensity response to tip-tilts in the presence of WFEs. 
We validate this behavior with experiments in \S\ref{ssec:tiptilt_WFE}.

\section{Experimental characterization of a 3-port PL on SCExAO}\label{sec:experiment}

In this section, we present our experimental characterization of the 3-port PL on the SCExAO instrument \cite{jovanovic15} at the Subaru Telescope. We first describe our experimental setup in \S\ref{ssec:setup}. In \S\ref{ssec:tiptilt}, we show tip-tilt sensitivity of the 3-port PL as a function of wavelength, $B_n (\lambda)$, and compare with predictions from \S\ref{ssec:chromaticity}. In \S\ref{ssec:tiptilt_WFE} we present intensity response to tip-tilt in presence of time-varying WFEs, $\langle B_n (\lambda)\rangle_t$, in connection to \S\ref{ssec:WFE}. In \S\ref{ssec:couplingmap}, we show measured coupling maps, the measured PL output spectra as a function of ($x$, $y$) fiber positions in the focal plane. In \S\ref{ssec:mapfitting} we present our initial results of modeling coupling maps (retrieving transfer matrix components) using the models presented in \S\ref{sec:model}.

\subsection{Experimental setup}\label{ssec:setup}

\begin{figure}
    \centering
    \includegraphics[width=1\linewidth]{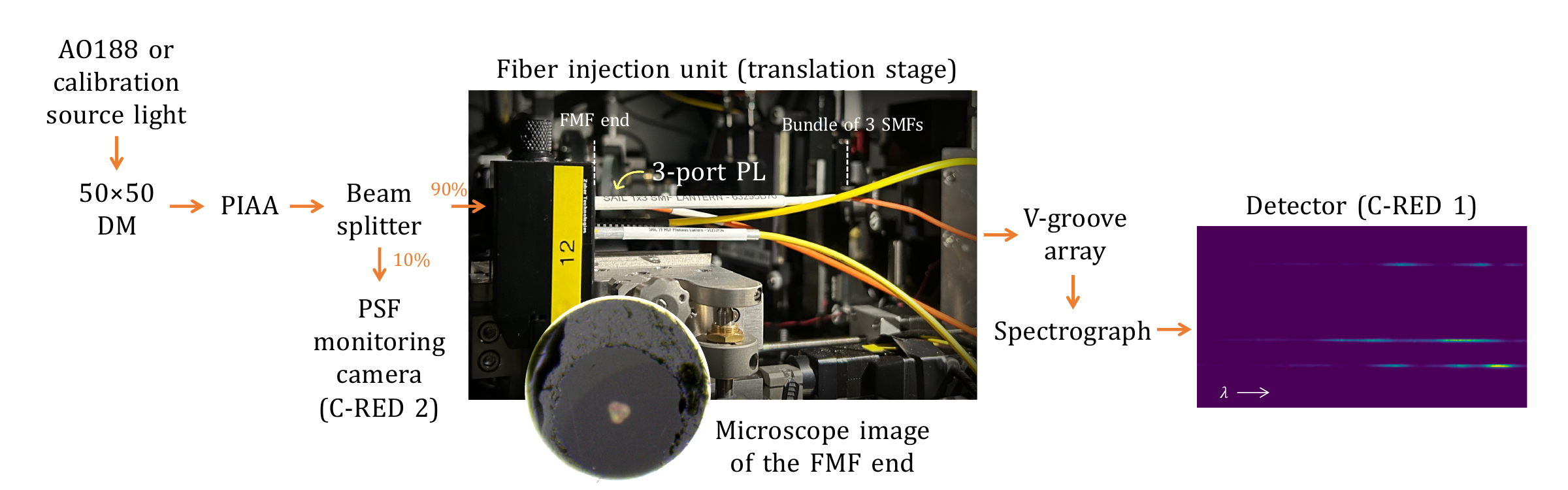}
    \caption{The instrument setup for experiments. The microscope image of the FMF end of the 3-port PL is shown in the inset picture, illuminated by retro-injecting visible light from one of the SMF outputs for visibility. }
    \label{fig:setup}
\end{figure}

The PL used in the experiment is a 3-port pigtailed PL, manufactured at Sydney Astrophotonics Instrumentation Laboratory. Three SMFs (SMF-28e) are arranged in an equilateral triangle without using a preform, placed in a lower-index capillary, then tapered down such that the cladding of the SMFs becomes a core and the capillary becomes the cladding in the FMF end. The microscope image of the FMF end illuminated by retro-injecting visible light from one of the SMF outputs is shown as an inset picture in Figure \ref{fig:setup}. The diameter of the core of the FMF end is 17.8~\textmu m, designed to support three modes in $H$ band. The average throughput over the three ports is measured to be 84\%. 

The 3-port PL is installed on the infrared bench of the SCExAO high-contrast testbed \cite{jovanovic15} at the Subaru Telescope. Figure \ref{fig:setup} illustrates the setup. Light is emitted from a supercontinuum laser, collimated by an off-axis parabolic (OAP) mirror, and then sent to a 2040-actuator deformable mirror (DM). Then the light is apodized passing through the phase-induced amplitude apodization (PIAA) lenses, making the beam shape Gaussian \cite{guyon03}. Then the beam is split by a 90:10 beam splitter, sending 10\% of the light to the PSF monitoring camera (First Light Imaging C-RED 2 camera). 90\% of the light is focused by an OAP and sent to the fiber injection unit \cite{jovanovic17} where the 3-port PL is mounted. The fiber injection unit consists of a translation stage which allows the fiber to move in $x$, $y$, and $f$, changing the position of the fiber in the focal plane transverse and parallel to the beam. In front of the fiber is a focusing lens. The lens and the translation stage are mounted on a carriage that can change the distance between the focusing lens and the OAP mirror, thus changing the focal ratio \cite{jovanovic17,lin22spie}. Therefore, by adjusting $x$, $y$, $f$, and the carriage position, we can optimize light injection into the FMF entrance of the PL.

After the fiber injection unit, the SMF outputs are spliced to a linear v-groove and sent to the spectrograph. The v-groove array is on a motorized x/y stage at the entrance of the spectrograph, to allow precise alignment of the spectral traces on the detector. A first achromatic lens collimates the beam onto the dispersive prism. The lens can be adjusted to focus the beams on the detector. The prism can be replaced by a mirror, to go from a spectroscopic mode to a photometric mode. Finally, a fixed focusing lens images the traces on the detector, a First Light Imaging C-RED ONE camera. The spectrograph was designed to be diffraction-limited from 1 to 1.8 $\mu$m, covering $y$, $J$ and $H$ bands, with a spectral resolution of $R \sim 700$, 500 and 300 in $y$, $J$ and $H$ respectively.

\begin{figure}[hbt!]
    \centering
    \includegraphics[width=1\linewidth]{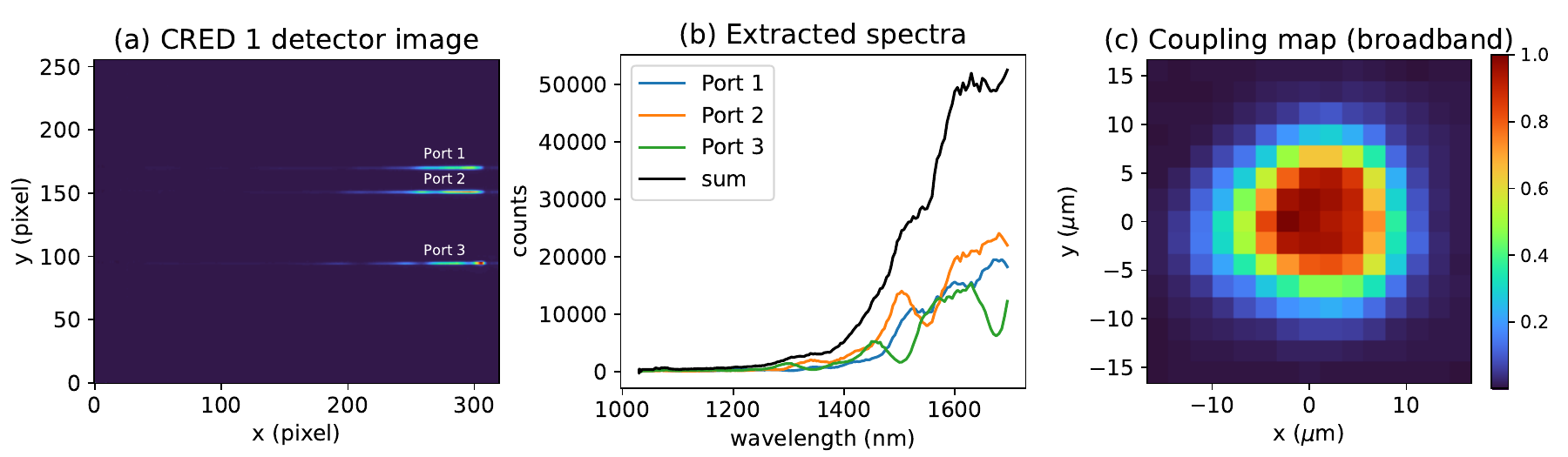}
    \caption{(Left) The example detector image of the three dispersed outputs. (Middle) The extracted spectra of individual traces and the summed spectra. (Right) The coupling map showing the total flux (stacked over all the wavelengths and the ports) taken by recording the flux while moving the $x$, $y$ fiber positions. The spectra in the left and middle panels correspond to the spectra at the origin of this map.}
    \label{fig:example_spectra}
\end{figure}

Figure \ref{fig:example_spectra} displays the example spectra of the supercontinuum laser source. The left panel shows the detector image, with the three spectral traces. The middle panel shows the extracted spectra, computed by simply summing up the pixels vertically. A tunable filter is used to find the wavelength solution, by adjusting the wavelength while recording the spots on the detector, to build a pixel-wavelength mapping model. The right panel shows the coupling map, the total flux at each $x$, $y$ fiber position with fixed $f$ and the carriage position. The spectra shown in the left and middle panels correspond to the origin of this map, which is close to the weighted centroid position of the map.

\subsection{Linear intensity response to tip-tilt}\label{ssec:tiptilt}

In this section, we present experimentally measured linear intensity responses to tip-tilts, the $B_n$ matrix. Having large linear intensity response corresponds to larger sensitivity to high angular resolutions (Paper I), and is related to the relative phase between \LP{01} and \LP{11} modes as discussed in \S\ref{ssec:chromaticity}. 

To measure wavelength-dependent intensity response to tip-tilts, we first optimize the injection into the 3-port PL by taking the coupling map and moving to the weighted centroid position. The normalized intensity in this position corresponds to $\textbf{I}_{n0}$. Next we sequentially apply tip-tilt aberrations using the DM while recording the spectra, in the range of [-0.4, 0.4] radians RMS (at a reference wavelength of 1.55~\textmu m), with steps of 0.1 radians RMS. Then we calculate normalized spectra by dividing the spectrum in each port by the summed spectrum. Then in each wavelength channel, the intensity response in each port as a function of the tip-tilt aberration amplitude is fitted by a second-order polynomial, $y= ax^2 + bx + c$. The slope $b$ then corresponds to the first-order response to the tip-tilt aberrations at the origin, the linear intensity response to tip-tilt ($B_n$). We repeat this procedure 10 times and average them. 

Top panels of Figure \ref{fig:tiptilt_response} show the linear intensity responses as a function of wavelength. As expected from the model and numerical simulations presented in Figure 9 of Paper I, qualitatively, the relative intensity responses exhibit sinusoidal behavior as a function of wavelength. 

\subsection{Linear intensity response to tip-tilt, with turbulence}\label{ssec:tiptilt_WFE}

\begin{figure}[hbt!]
    \centering
    \includegraphics[width=1\linewidth]{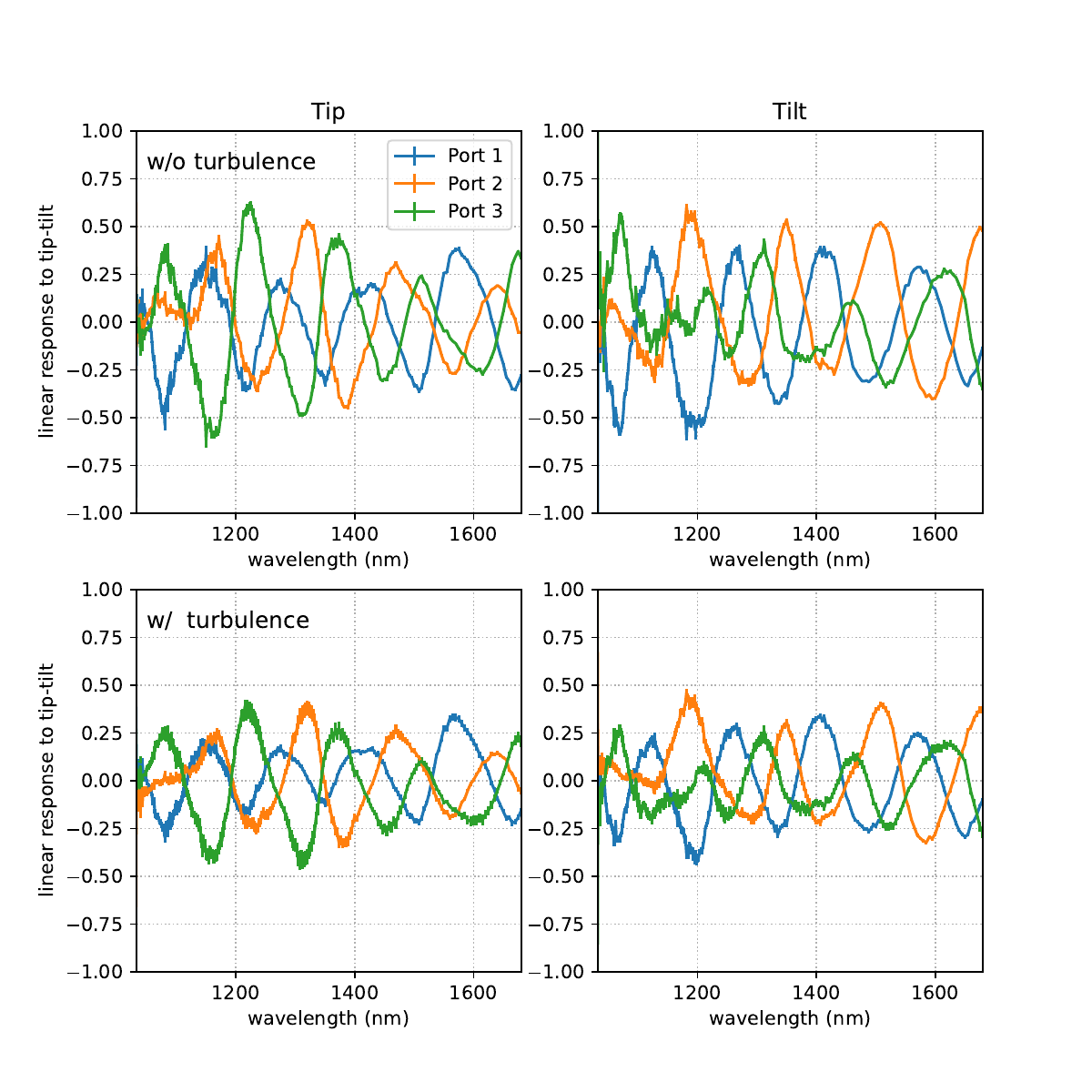}
    \caption{(Top) Measured first-order intensity response to tip-tilts (the $B_n$ matrix, with units of radians RMS$^{-1}$) as a function of wavelength. (Bottom) The same figure but with turbulence of 0.15~\textmu m. The responses are systematically decreased compared to the measurements without turbulence.}
    \label{fig:tiptilt_response}
\end{figure}

In \S\ref{ssec:WFE} we discussed the systematic decrease of intensity response to tip-tilts in a long exposure regime with time-varying turbulence. An experimental demonstration of this effect is shown in bottom panels of Figure \ref{fig:tiptilt_response}. The time-varying phase aberrations with amplitude of 0.15~\textmu m was simulated on the SCExAO real-time computer and applied on the DM.
With the simulated turbulence running on the DM, we took the tip-tilt scan five times each as described above. We repeated this with three realizations of turbulence, resulting in 15 measurements in total.
The sinusoidal behavior of the tip-tilt responses remains unchanged but the amplitudes are decreased, as expected from \S\ref{ssec:WFE}. 

In real spectroastrometric observations, such systematic effects should be calibrated in order not to underestimate the centroid shifts. 
This differs from standard spectroastrometry, which measures centroid shifts directly from the focal plane image. While WFEs broaden the PSF and degrade the S/N of the centroid shift measurement, they do not bias the centroid shift itself, although additional complications may exist introduced by slits or focal plane resampling.
We defer developing practical calibration strategies to future work.

\subsection{Coupling maps as a function of wavelength}\label{ssec:couplingmap}

In Equation \ref{eq:linear}, $\textbf{I}_{n0}$ and $B_n$ are defined for an on-axis reference PSF. 
However, in real observations, the average position of the PSF ($x$, $y$) over an exposure may deviate from the on-axis position.
Thus, it is useful to have a map of $\textbf{I}_{n}$ and $B_n$ for the ($x$, $y$) fiber positions in the focal plane to correct for such pointing errors.
These maps can be inferred from coupling maps, 
which are spectral responses as a function of ($x$, $y$) fiber position in the focal plane.

\begin{figure}
    \centering
    \includegraphics[width=0.6\linewidth]{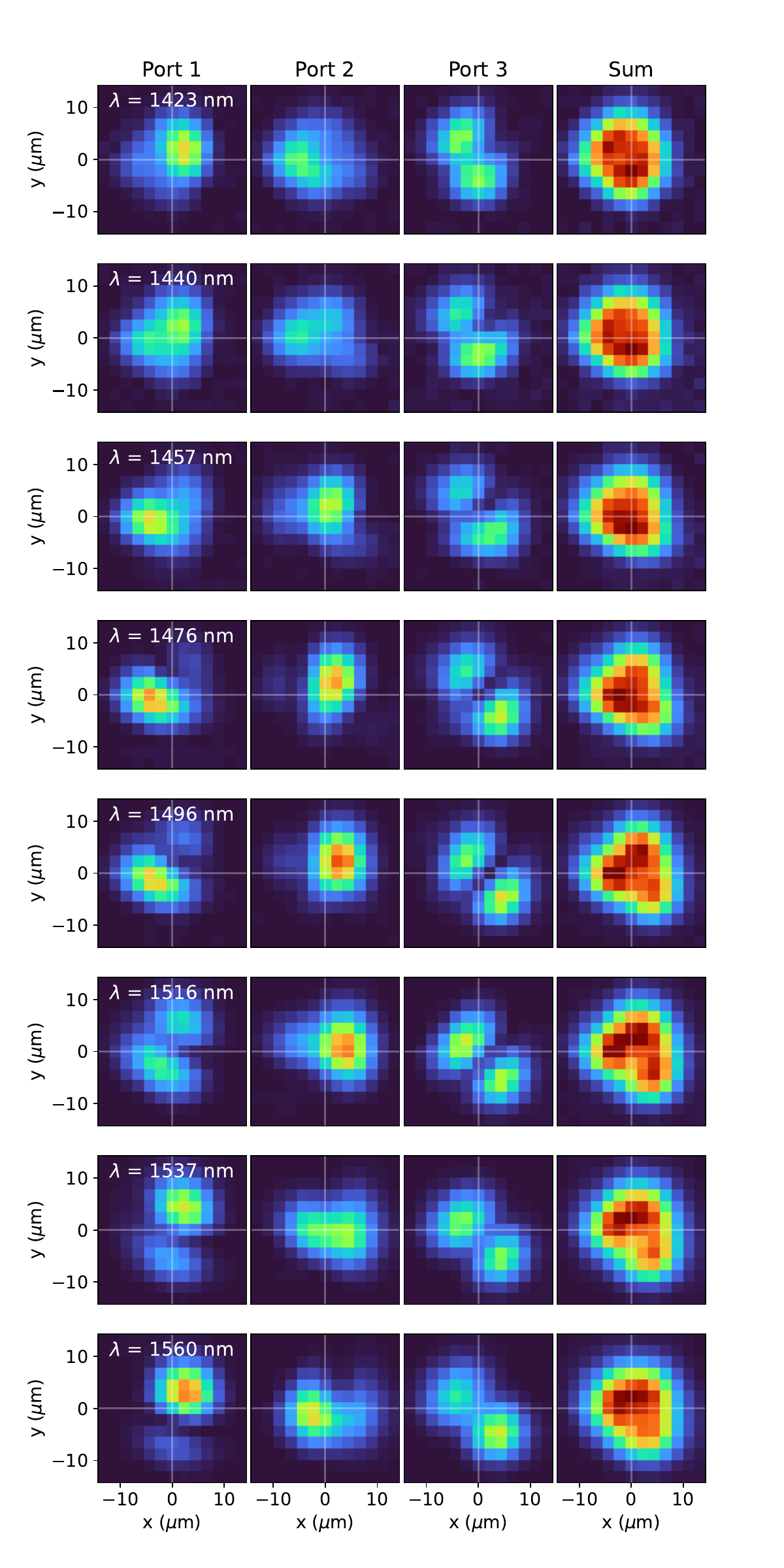}
    \caption{Measured coupling maps of individual ports and the sum for a few wavelengths.}
    \label{fig:coupling_maps_many}
\end{figure}

Figure \ref{fig:coupling_maps_many} shows the measured coupling maps of individual ports at a few wavelengths. The right panels are the summed coupling maps of the three individual maps. The individual maps divided by the summed map correspond to the $\textbf{I}_{n}$ map at any ($x$, $y$) position, and its gradients along the $x$, $y$ directions correspond to $B_n$ map. Thus, in the presence of time-varying pointing errors, in every frame, the observed normalized spectra at the reference wavelengths (where zero spectroastrometric signals are expected) can be fit to previously measured coupling maps $\textbf{I}_{n}(x,y; \lambda)$ to find the pointing error ($x$, $y$) and the corresponding $B_n$ at the position. 

\begin{figure}
    \centering
    \includegraphics[width=0.85\linewidth]{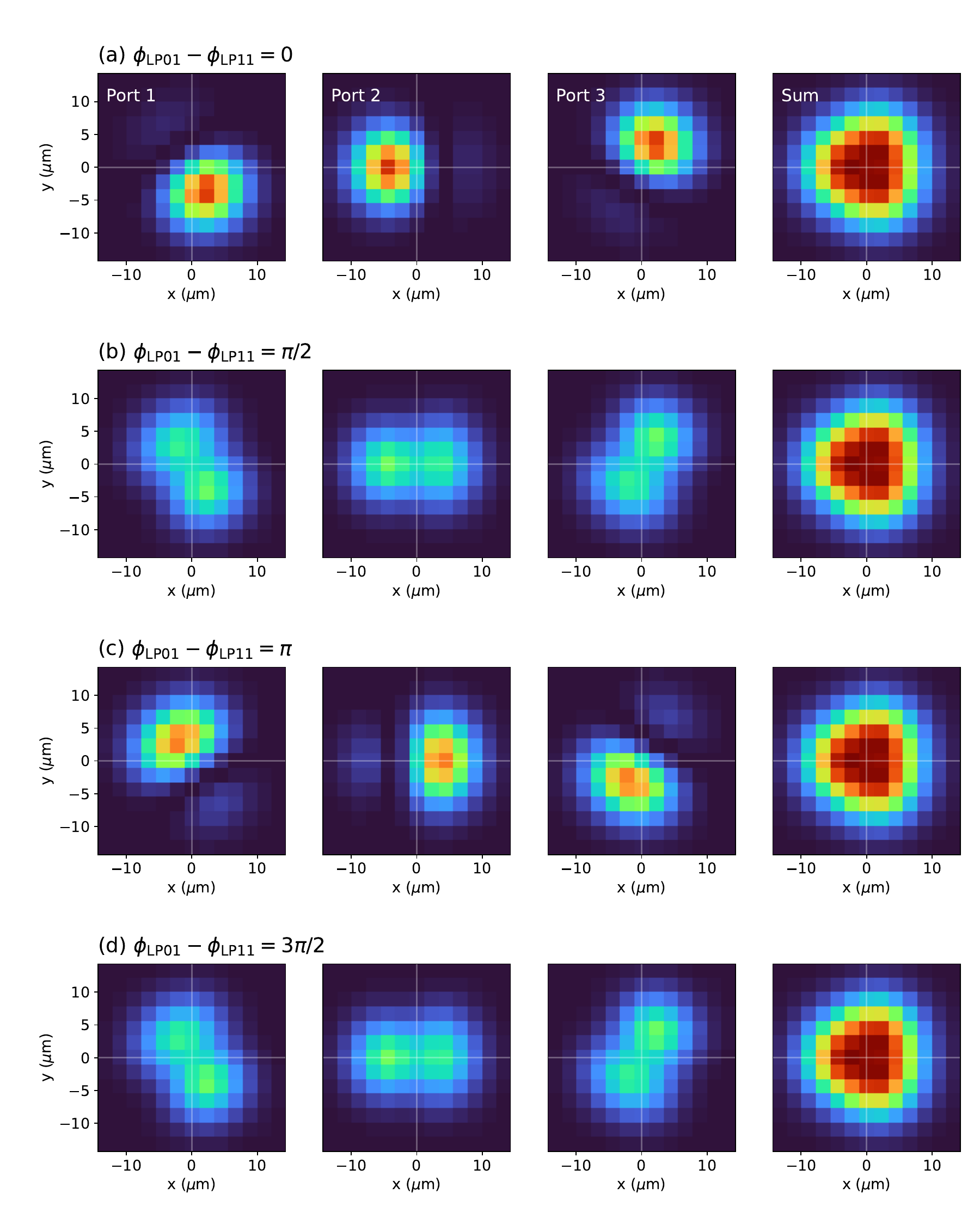}
    \caption{Simulated coupling maps of individual ports and the sum assuming a standard 3-port PL and wavelength of $\lambda = 1550$~nm for four cases: phase difference between \LP{01} and \LP{11} is (a) zero, (b) $\pi/2$, (c) $\pi$, and (d) $3\pi/2$. Since the majority of the wavelength dependence in PL responses arises from phase differences between the \LP{01} and \LP{11} modes, this simulation effectively captures the wavelength dependence of the measured coupling maps shown in Figure \ref{fig:coupling_maps_many}. Note that (a) and (c) correspond to the case of the greatest linear intensity response to tilt (\S\ref{ssec:chromaticity}), with greatest gradients at the origin of the individual coupling maps. Conversely, (b) and (d) correspond to the case of nonlinear intensity response to the tilt; the gradients at the origin of the individual maps are zero and the symmetric coupling maps indicate the ambiguity of recovering tip-tilts from lantern intensity responses even with larger tip-tilt displacements.}
    \label{fig:coupling_maps_simulated}
\end{figure}

Here we describe a few features in the coupling maps and compare with numerical simulations. The coupling maps of individual ports peak at different $x$, $y$ fiber positions, with characteristic double blob shapes. Also, the coupling maps gradually change with wavelength. The most prominent feature of wavelength dependence is the alternating peaks between the blobs. This is due to the wavelength dependence of the \LP{01} and \LP{11} phase differences in the transfer matrix, as simulated in Figure \ref{fig:coupling_maps_simulated} for a standard 3-port PL. Besides this feature, we find that the blobs rotate with wavelength. This may be explained by the chromatic behavior of the amplitude components of the transfer matrix. Note that large gradients in the coupling maps indicate better sensitivity to tip-tilts. We also find that the peak of the summed coupling map (right panels) as well as its overall position moves with wavelength. The shifting peak is expected to be caused by wavelength-dependent losses within the lantern. However, the movement of the overall position as a function of wavelength is of unknown origin at the moment and we leave this for future investigation.

\subsection{Measuring the transfer matrix using coupling maps}\label{ssec:mapfitting}

In \S\ref{sec:model}, we described a method of modeling PL intensity responses. If the transfer matrices $\tilde{\eta}_{ij}$s for every wavelength channel are determined, one can build the PL spectral response models for any input scene. One way of measuring $\tilde{\eta}_{ij}$s as a function of wavelength is using the coupling maps of individual ports as a function of wavelength. Moving the injection stage in $x$ and $y$ corresponds to the tip-tilt aberration in the pupil plane, modulating the $\tilde{W}_{jk}$ components while leaving transfer matrix components unchanged. Additional phase probes may be applied using the DM to achieve more diverse $\tilde{W}_{jk}$s. Using Equation \ref{eq:overlap}, $\tilde{W}_{jk}$ can be computed on each $x$, $y$ fiber position with appropriate models of the pupil and the fiber entrance modes. Then with the measured intensities $I_i$ at various $x$, $y$ positions (and potentially with additional phase probes using the DM) and corresponding $\tilde{W}_{jk}$s, the transfer matrix components $\tilde{\eta_{ij}}$ can be solved except for the piston phases (e.g., \LP{01} phases for each port) using Equation \ref{eq:intensity1}.

\begin{figure}
    \centering
    \includegraphics[width=1\linewidth]{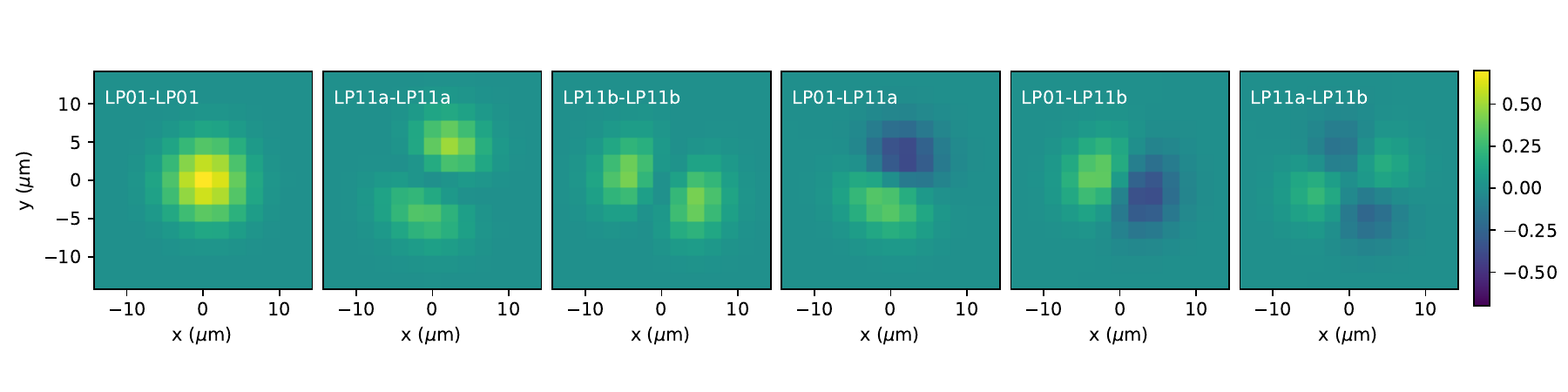}
    \caption{The six components that constitute the coupling maps of a 3-port PL. }
    \label{fig:overlap_components}
\end{figure}

Given that our system has PIAA beam-shaping lenses, we simplify the problem by assuming that the PSF is purely flat-phase Gaussian profile beam and rewriting the models in the focal plane instead of the pupil plane. While applying tip-tilt phase probes on the DM changes the amplitude and phase profile of the PSF with the PIAA lenses, fiber injection stage movement does not change the PSF itself. Writing the focal plane field at the fiber positions ($x,y$) as $E_f (x,y)$, the intensity in the $i$-th port is:
\begin{align}\label{eq:inten_focalplane}
    \begin{aligned}
        I_i (x,y) = |\tilde{E}_i (x,y)|^2 &= \left|\sum_{j=1}^{N} \int\tilde{\eta}_{ij} E_f(x,y) X_{f,j} dA\right|^2 \\ &= \sum_{j=1}^{N} \eta_{ij}^2 X'^2_j(x,y) 
        + \sum_{j=k+1}^{N}\sum_{k=1}^{N} 2\eta_{ij}\eta_{ik} \cos{(\phi_{ik}-\phi_{ij})} X'_j(x,y) X'_k(x,y)
    \end{aligned}
\end{align}
where
\begin{equation}
    X'_j(x,y) \equiv \int E_f(x,y) X_{f,j} dA.
\end{equation}
Note that the integral is evaluated over the focal plane. With a model of $X_{f,j}$ (panel (a) of Figure \ref{fig:modes}) and the PSF, $X'_j(x,y) X'_k(x,y)$ can be computed on a grid of ($x,y$). Figure \ref{fig:overlap_components} shows the six maps of $X'_j(x,y) X'_k(x,y)$. The coupling map of the $i$-th port can then be described as a linear combination of these six components with the six coefficients ($\eta_{ij}^2$ and $2\eta_{ij}\eta_{ik}\cos{(\phi_{ik}-\phi_{ij})}$). An iterative approach is used to fit the coupling maps: finding PSF and fiber entrance models that minimize the squared residuals with respect to the coupling maps. We fix the shapes of the PSF as Gaussian and the fiber entrance modes as the modes solved for our 3-port PL, but allow the following parameters to vary in our fitting procedure: fiber orientation angle, $x$,$y$ fiber centroid positions, and the PSF size.

\begin{figure}
    \centering
    \includegraphics[width=1\linewidth]{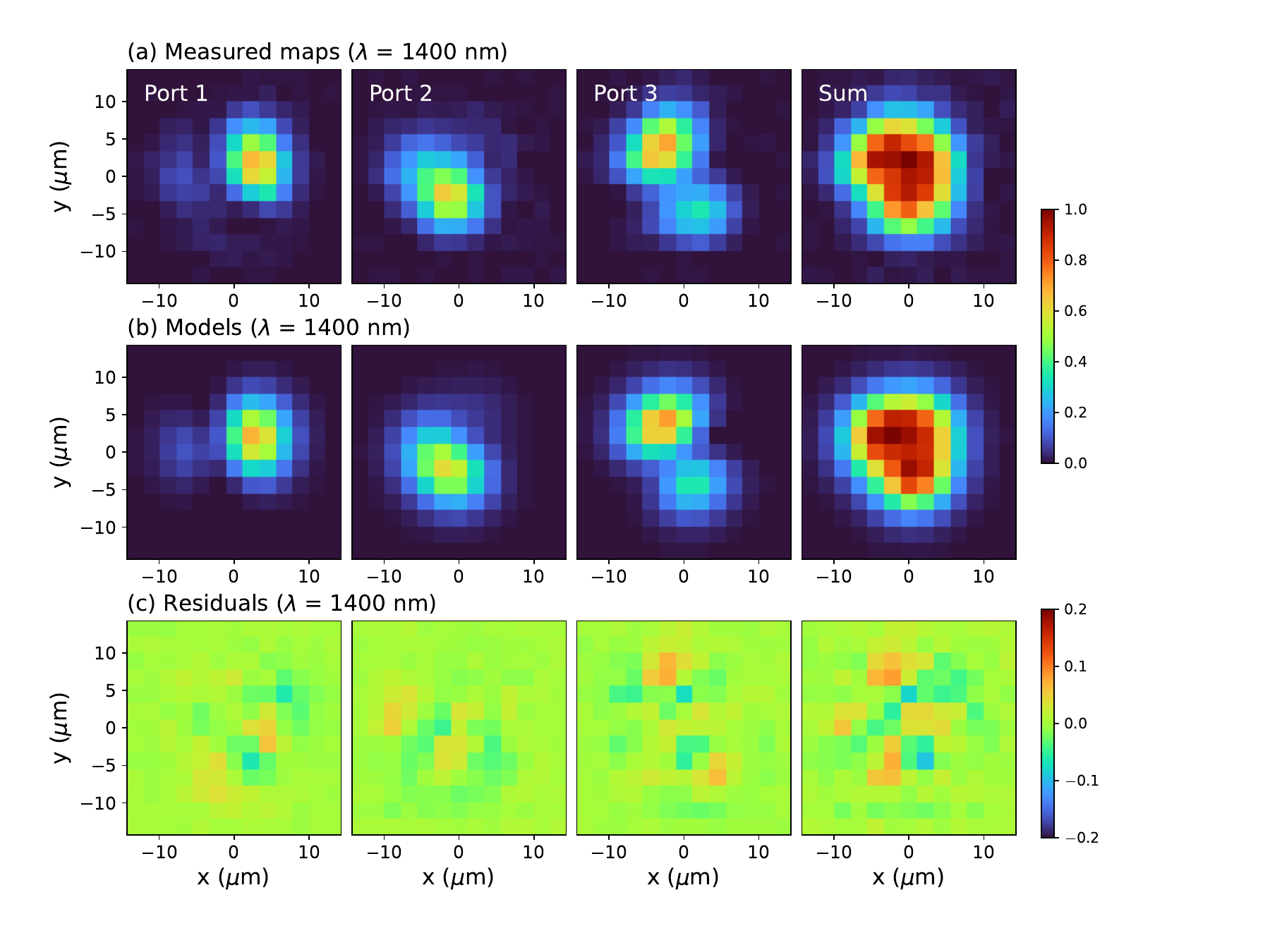}
    \caption{(a) Measured coupling maps ($\lambda = 1400$~nm) of individual ports and their sum, (b) modeled coupling maps, as linear combination of overlap components shown in Figure \ref{fig:overlap_components}, and (c) residuals. }
    \label{fig:couplingmap_lam1}
\end{figure}

\begin{figure}
    \centering
    \includegraphics[width=1\linewidth]{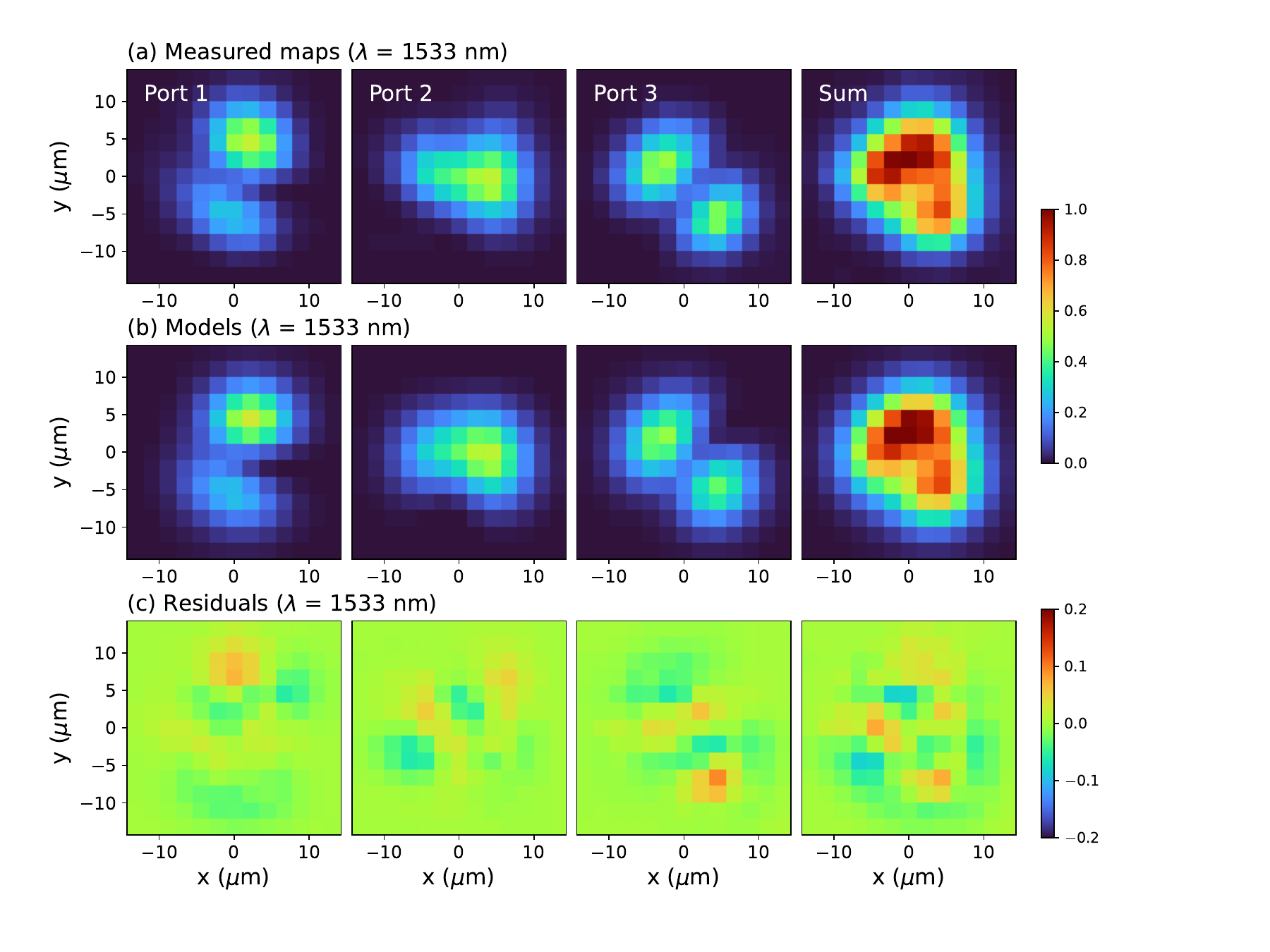}
    \caption{Same as Figure \ref{fig:couplingmap_lam1} but at $\lambda = 1533$~nm.}
    \label{fig:couplingmap_lam2}
\end{figure}

Figures \ref{fig:couplingmap_lam1} and \ref{fig:couplingmap_lam2} show the results of the coupling map modeling for two wavelengths, $\lambda = 1400$~nm and $\lambda = 1533$~nm, respectively. The models capture most of the features in the coupling maps. We compute the dot product of the measured coupling maps and the fitted models as a metric to assess the model fit, which gives 0.992 and 0.994 for $\lambda = 1400$~nm and $\lambda = 1533$~nm, respectively. The coupling map models are relatively stable across time; we fit the coupling maps taken two days later using the coefficients retrieved from the coupling map fitting shown in Figures \ref{fig:couplingmap_lam1} and \ref{fig:couplingmap_lam2}, obtaining dot products of 0.929 and 0.960 for the two wavelengths. Although the dot products are lower in this case, which may be due to the drift of the spectral traces on the detector in wavelength, we find the morphology of the coupling maps and their wavelength-dependent behavior are consistent over time.

The residuals of the model fit exhibit systematic patterns rather than random fluctuations. This could be explained by the departure of the PSF during the exposure to the ideal Gaussian PSF model, in particular due to the bench jitter. In order to reliably measure the transfer matrix components, the bench jitter along with any other misalignments or aberrations need to be carefully modeled. Moreover, the model fitting accuracy can be improved by jointly modeling the coupling maps across multiple wavelength channels. We defer such refinements to future work.

\section{Conclusion and future work}\label{sec:conclusion}

Spectroastrometric signals can be measured with PLs from the relative spectra between the ports (Paper I). For spectroastrometric measurements, the sensitivity of PL intensity response to tip-tilt is a crucial factor, which depends on wavelength and needs to be characterized. To this end, we have developed PL spectral response models in \S\ref{sec:model} and showed initial results of experimental spectral characterization of the 3-port PL on the SCExAO test bench. Paper I expected that intensity response to tip-tilts as a function of wavelength would exhibit sinusoidal behavior for a 3-port PL, and the responses would be modified with WFE averaging. We have verified these behaviors with experiments, in \S\ref{sec:experiment}. In addition, we presented coupling maps as a function of wavelength and discussed how they would be useful for offsetting pointing errors. Furthermore, we showed initial results on recovering PL transfer matrices using the coupling maps and the analytical framework illustrated in \S\ref{sec:model}. We acknowledge that our transfer matrix recovery approach shares some similarity with the approach described by Ref. \cite{sengupta24} and note that our research was conducted independently.

In \S\ref{ssec:tiptilt_WFE} we discussed the neccessity of practical calibration strategies for time-varying WFEs to accurately measure spectroastrometric signals. There are two directions for future work aimed at achieving this goal. The first involves building a reliable PL spectral response model and developing methods to leverage this model for extracting and calibrating spectroastrometric signals from measured spectra. This requires
refining the transfer matrix recovery procedure, by relaxing the Gaussian PSF assumption and accounting for the PIAA optics, misalignments, or aberrations that may present in the system. Additionally, more diverse probes can be used by applying phase probes on the DM in addition to the fiber injection stage movement. Furthermore, data acquisition and fitting strategies can be improved. It may also be necessary to consider the two polarization states in the model. The stability of recovered transfer matrices over time needs to be studied as well.

The other direction is to take a data-driven approach, not relying on the models. The spectral channels outside the spectral feature of interest for spectroastrometry can be used for wavefront sensing, both identifying systematic tip-tilts and the WFE averaging effects. Also, the calibration strategies may be augmented by specific observation strategies, for instance 
by taking on-sky coupling maps, by observing a calibrator star taken with similar WFE properties, by simultaneously recording the PSF with the PSF monitoring camera, using AO telemetry data, or by dithering the fiber position on the target to measure the $B_n$ matrix directly. 
We plan to develop practical calibration and observation strategies and test on-sky spectroastrometric measurement with the 3-port PL in the future.

\subsection* {Code, Data, and Materials Availability} 

The data and code used in preparation of this work are available on GitHub (\url{https://github.com/YooJung-Kim/3PL_characterization/}).

\subsection* {Acknowledgments}
This paper is based on the experiments previously reported in SPIE proceedings, Ref. \cite{kim24SPIE}.
This work is supported by the National Science Foundation under Grant Nos. 2109231, 2109232, 2308360, and 2308361. 
The development of SCExAO is supported by the Japan Society for the Promotion of Science (Grant-in-Aid for Research  No. 23340051, 26220704, 23103002, 19H00703, 19H00695 and 21H04998), the Subaru Telescope, the National Astronomical Observatory of Japan, the Astrobiology Center of the National Institutes of Natural Sciences, Japan, the Mt Cuba Foundation and the Heising-Simons Foundation. The authors wish to recognize and acknowledge the very significant cultural role and reverence that the summit of Maunakea has always had within the indigenous Hawaiian community, and are most fortunate to have the opportunity to conduct observations from this mountain.


\bibliography{report}   

\begin{thebibliography}{10}

\bibitem{bailey98}
J.~A. {Bailey}, ``{Spectroastrometry: a new approach to astronomy on small spatial scales},'' in {\em Optical Astronomical Instrumentation},  S.~{D'Odorico}, Ed., {\em Society of Photo-Optical Instrumentation Engineers (SPIE) Conference Series} {\bf 3355}, 932--939  (1998).

\bibitem{whelan08}
E.~{Whelan} and P.~{Garcia}, ``{Spectro-astrometry: The Method, its Limitations, and Applications},'' in {\em Jets from Young Stars II},  F.~{Bacciotti}, L.~{Testi}, and E.~{Whelan}, Eds.,  {\bf 742}, 123  (2008).

\bibitem{gravity_yso_i}
{GRAVITY Collaboration}, K.~{Perraut}, L.~{Labadie}, {\em et~al.}, ``{The GRAVITY Young Stellar Object survey. I. Probing the disks of Herbig Ae/Be stars in terrestrial orbits},'' {\em \aap} {\bf 632}, A53  (2019).

\bibitem{pontoppidan08}
K.~M. {Pontoppidan}, G.~A. {Blake}, E.~F. {van Dishoeck}, {\em et~al.}, ``{Spectroastrometric Imaging of Molecular Gas within Protoplanetary Disk Gaps},'' {\em \apj} {\bf 684}, 1323--1329  (2008).

\bibitem{pontoppidan11}
K.~M. {Pontoppidan}, G.~A. {Blake}, and A.~{Smette}, ``{The Structure and Dynamics of Molecular Gas in Planet-forming Zones: A CRIRES Spectro-astrometric Survey},'' {\em \apj} {\bf 733}, 84  (2011).

\bibitem{brittain15}
S.~D. {Brittain}, J.~R. {Najita}, and J.~S. {Carr}, ``{Near infrared high resolution spectroscopy and spectro-astrometry of gas in disks around Herbig Ae/Be stars},'' {\em \apss} {\bf 357}, 54  (2015).

\bibitem{bosco21}
F.~{Bosco}, J.~F. {Hennawi}, J.~{Stern}, {\em et~al.}, ``{Spatially Resolving the Kinematics of the {\ensuremath{\lesssim}}100 {\ensuremath{\mu}}as Quasar Broad-line Region Using Spectroastrometry. II. The First Tentative Detection in a Luminous Quasar at z = 2.3},'' {\em \apj} {\bf 919}, 31  (2021).

\bibitem{gravity18}
{Gravity Collaboration}, E.~{Sturm}, J.~{Dexter}, {\em et~al.}, ``{Spatially resolved rotation of the broad-line region of a quasar at sub-parsec scale},'' {\em \nat} {\bf 563}, 657--660  (2018).

\bibitem{gravity_yso_vii}
{GRAVITY Collaboration}, V.~{Ganci}, L.~{Labadie}, {\em et~al.}, ``{The GRAVITY young stellar object survey. VIII. Gas and dust faint inner rings in the hybrid disk of HD141569},'' {\em \aap} {\bf 655}, A112  (2021).

\bibitem{gravity_yso_x}
{Gravity Collaboration}, A.~{Soulain}, K.~{Perraut}, {\em et~al.}, ``{The GRAVITY young stellar object survey. X. Probing the inner disk and magnetospheric accretion region of CI Tau},'' {\em \aap} {\bf 674}, A203  (2023).

\bibitem{bra06}
E.~{Brannigan}, M.~{Takami}, A.~{Chrysostomou}, {\em et~al.}, ``{On the detection of artefacts in spectro-astrometry},'' {\em \mnras} {\bf 367}, 315--322  (2006).

\bibitem{whelan15}
E.~T. {Whelan}, N.~{Hu{\'e}lamo}, J.~M. {Alcal{\'a}}, {\em et~al.}, ``{Spectro-astrometry of LkCa 15 with X-Shooter: Searching for emission from LkCa 15b},'' {\em \aap} {\bf 579}, A48  (2015).

\bibitem{dav10}
B.~{Davies}, S.~L. {Lumsden}, M.~G. {Hoare}, {\em et~al.}, ``{The circumstellar disc, envelope and bipolar outflow of the massive young stellar object W33A},'' {\em \mnras} {\bf 402}, 1504--1515  (2010).

\bibitem{got12}
M.~{Goto}, A.~{Carmona}, H.~{Linz}, {\em et~al.}, ``{Kinematics of Ionized Gas at 0.01 AU of TW Hya},'' {\em \apj} {\bf 748}, 6  (2012).

\bibitem{mur13}
K.~{Murakawa}, S.~L. {Lumsden}, R.~D. {Oudmaijer}, {\em et~al.}, ``{Near-infrared integral field spectroscopy of massive young stellar objects},'' {\em \mnras} {\bf 436}, 511--525  (2013).

\bibitem{kim24JATIS}
Y.~J. {Kim}, M.~P. {Fitzgerald}, J.~{Lin}, {\em et~al.}, ``{On the potential of spectroastrometry with photonic lanterns},'' {\em Journal of Astronomical Telescopes, Instruments, and Systems}   (submitted).

\bibitem{leo10}
S.~G. {Leon-Saval}, A.~{Argyros}, and J.~{Bland-Hawthorn}, ``{Photonic lanterns: a study of light propagation in multimode to single-mode converters},'' {\em Optics Express} {\bf 18}, 8430  (2010).

\bibitem{leo13}
S.~G. {Leon-Saval}, A.~{Argyros}, and J.~{Bland-Hawthorn}, ``{Photonic lanterns},'' {\em Nanophotonics} {\bf 2}, 429--440  (2013).

\bibitem{bir15}
T.~A. {Birks}, I.~{Gris-S{\'a}nchez}, S.~{Yerolatsitis}, {\em et~al.}, ``{The photonic lantern},'' {\em Advances in Optics and Photonics} {\bf 7}, 107  (2015).

\bibitem{cor18}
M.~K. {Corrigan}, T.~J. {Morris}, R.~J. {Harris}, {\em et~al.}, ``{Demonstration of a photonic lantern low order wavefront sensor using an adaptive optics testbed},'' in {\em Adaptive Optics Systems VI},  L.~M. {Close}, L.~{Schreiber}, and D.~{Schmidt}, Eds., {\em Society of Photo-Optical Instrumentation Engineers (SPIE) Conference Series} {\bf 10703}, 107035H  (2018).

\bibitem{nor20}
B.~R.~M. {Norris}, J.~{Wei}, C.~H. {Betters}, {\em et~al.}, ``{An all-photonic focal-plane wavefront sensor},'' {\em Nature Communications} {\bf 11}, 5335  (2020).

\bibitem{cruz-delgado21}
D.~Cruz-Delgado, J.~C. Alvarado-Zacarias, M.~A. Cooper, {\em et~al.}, ``Photonic lantern tip/tilt detector for adaptive optics systems,'' {\em Opt. Lett.} {\bf 46}, 3292--3295  (2021).

\bibitem{lin22}
J.~{Lin}, M.~P. {Fitzgerald}, Y.~{Xin}, {\em et~al.}, ``{Focal-plane wavefront sensing with photonic lanterns: theoretical framework},'' {\em Journal of the Optical Society of America B Optical Physics} {\bf 39}, 2643  (2022).

\bibitem{lin23}
J.~{Lin}, M.~P. {Fitzgerald}, Y.~{Xin}, {\em et~al.}, ``{Focal-plane wavefront sensing with photonic lanterns II: numerical characterization and optimization},'' {\em Journal of the Optical Society of America B Optical Physics} {\bf 40}, 3196  (2023).

\bibitem{lin23b}
J.~W. {Lin}, M.~P. {Fitzgerald}, Y.~{Xin}, {\em et~al.}, ``{Real-time Experimental Demonstrations of a Photonic Lantern Wave-front Sensor},'' {\em \apjl} {\bf 959}, L34  (2023).

\bibitem{lin21}
J.~{Lin}, N.~{Jovanovic}, and M.~P. {Fitzgerald}, ``{Design considerations of photonic lanterns for diffraction-limited spectrometry},'' {\em Journal of the Optical Society of America B Optical Physics} {\bf 38}, A51  (2021).

\bibitem{vievard24}
S.~{Vievard}, M.~{Lallement}, S.~{Leon-Saval}, {\em et~al.}, ``{Spectroscopy using a visible photonic lantern at the Subaru telescope: Laboratory characterization and first on-sky demonstration on Ikiiki ($\alpha$ Leo) and `Aua ($\alpha$ Ori)},'' {\em arXiv e-prints} , arXiv:2409.06958  (2024).

\bibitem{jovanovic15}
N.~{Jovanovic}, F.~{Martinache}, O.~{Guyon}, {\em et~al.}, ``{The Subaru Coronagraphic Extreme Adaptive Optics System: Enabling High-Contrast Imaging on Solar-System Scales},'' {\em \pasp} {\bf 127}, 890  (2015).

\bibitem{lin22spie}
J.~{Lin}, S.~{Vievard}, N.~{Jovanovic}, {\em et~al.}, ``{Experimental measurements of AO-fed photonic lantern coupling efficiencies},'' in {\em Society of Photo-Optical Instrumentation Engineers (SPIE) Conference Series},  {\em Society of Photo-Optical Instrumentation Engineers (SPIE) Conference Series} {\bf 12188}, 121882E  (2022).

\bibitem{xin24}
Y.~Xin, D.~Echeverri, N.~Jovanovic, {\em et~al.}, ``{Laboratory demonstration of a Photonic Lantern Nuller in monochromatic and broadband light},'' {\em Journal of Astronomical Telescopes, Instruments, and Systems} {\bf 10}(2), 025001  (2024).

\bibitem{kim24}
Y.~J. {Kim}, M.~P. {Fitzgerald}, J.~{Lin}, {\em et~al.}, ``{Coherent Imaging with Photonic Lanterns},'' {\em \apj} {\bf 964}, 113  (2024).

\bibitem{roddier81}
F.~{Roddier}, ``{The effects of atmospheric turbulence in optical astronomy},'' {\em Progess in Optics} {\bf 19}, 281--376  (1981).

\bibitem{guyon03}
O.~{Guyon}, ``{Phase-induced amplitude apodization of telescope pupils for extrasolar terrestrial planet imaging},'' {\em \aap} {\bf 404}, 379--387  (2003).

\bibitem{jovanovic17}
N.~{Jovanovic}, C.~{Schwab}, O.~{Guyon}, {\em et~al.}, ``{Efficient injection from large telescopes into single-mode fibres: Enabling the era of ultra-precision astronomy},'' {\em \aap} {\bf 604}, A122  (2017).

\bibitem{sengupta24}
A.~R. {Sengupta}, J.~{Diaz}, B.~L. {Gerard}, {\em et~al.}, ``{Photonic lantern wavefront reconstruction in a multi-wavefront sensor single-conjugate adaptive optics system},'' {\em arXiv e-prints} , arXiv:2406.07771  (2024).

\bibitem{kim24SPIE}
Y.~J. {Kim}, M.~P. {Fitzgerald}, J.~{Lin}, {\em et~al.}, ``{Spectral characterization of 3-port photonic lantern for spectroastrometry},'' {\em Society of Photo-Optical Instrumentation Engineers (SPIE) Conference Series} {\bf 13095}, 1309529  (2024).

\end{thebibliography}
\bibliographystyle{spiejour}   



\vspace{1ex}



\end{document}